\documentclass[prd,twocolumn,twoside,preprintnumbers,superscriptaddress,nofootinbib,showpacs,showkeys]{revtex4}
\usepackage{amsmath,slashed}
\usepackage{graphicx,graphics}
\usepackage{dcolumn}
\usepackage{color}
\usepackage[hyperfootnotes=false]{hyperref}
\usepackage{breakurl}
\usepackage{xspace}

\newcommand{\MeV}{\,\text{MeV}}
\newcommand{\GeV}{\,\text{GeV}}

\newcommand{\Br}{\text{Br}}

\newcommand{\diff}{\text{d}}
\newcommand{\eps}{\epsilon}
\newcommand{\mk}{M_K}
\newcommand{\mpi}{M_\pi}
\renewcommand{\Im}{\text{Im}\,}
\renewcommand{\Re}{\text{Re}\,}

\allowdisplaybreaks

\begin{document}
\preprint{CERN-TH-2016-001, INT-PUB-16-001, PSI-PR-16-001}

\title{Lepton flavor (universality) violation in rare kaon decays}

\author{Andreas Crivellin}
\affiliation{Theoretical Physics Department, CERN, CH--1211 Geneva 23, Switzerland}
\affiliation{Paul Scherrer Institut, CH--5232 Villigen PSI, Switzerland}
\author{Giancarlo D'Ambrosio}
\affiliation{INFN-Sezione di Napoli, Via Cintia, I--80126 Napoli, Italy}
\author{Martin Hoferichter}
\affiliation{Institute for Nuclear Theory, University of Washington, Seattle, Washington 98195-1550, USA}
\author{Lewis C.\ Tunstall}
\affiliation{Albert Einstein Center for Fundamental Physics,
Institute for Theoretical Physics,\\ University of Bern, Sidlerstrasse 5, CH--3012 Bern, Switzerland}

\begin{abstract}
Recent anomalies in the decays of $B$ mesons and the Higgs boson provide hints towards lepton flavor (universality) violating physics beyond the Standard Model. We observe that four-fermion operators which can explain the $B$-physics anomalies have corresponding analogs in the kaon sector, and we analyze their impact on $K\to\pi\ell \ell'$ and $K\to\ell \ell'$ decays $(\ell=\mu,e)$. For these processes, we note the corresponding physics opportunities at the NA62 experiment.
In particular, assuming minimal flavor violation, we comment on the required improvements in sensitivity necessary to test the $B$-physics anomalies in the kaon sector.  
\end{abstract}



\maketitle

\section{Puzzles in the flavor sector}
\label{intro}

The discovery of a Higgs-like resonance at the LHC experiments~\cite{Aad:2012tfa,Chatrchyan:2012ufa} 
provides the final ingredient to complete the Standard Model (SM) of particle physics. However, there are a variety of theoretical and phenomenological reasons to suspect that the SM is not the final theory, and that some form of new physics (NP) may also be present near the electroweak scale.  While no direct evidence for physics beyond the SM was found during the first LHC run, there are some interesting indirect hints for NP in the flavor sector, chiefly in the semileptonic decays of $B$ mesons and the SM-forbidden decay $h\to\mu\tau$ of the Higgs boson.

More specifically, deviations from the SM found by LHCb~\cite{Aaij:2013qta,Aaij:2015oid} in the decay $B\to K^* \mu^+\mu^-$ arise mainly in an angular observable called $P_5^\prime$~\cite{Descotes-Genon:2013vna}, with a significance of $2$--$3\sigma$ depending on assumptions made for the hadronic uncertainties~\cite{Descotes-Genon:2014uoa,Altmannshofer:2014rta,Jager:2014rwa}. In the decay $B_s\to\phi\mu^+\mu^-$, LHCb also uncovered~\cite{Aaij:2015esa} deviations compared to the SM prediction from lattice QCD~\cite{Horgan:2013pva,Horgan:2015vla} of $3.5\sigma$ significance~\cite{Altmannshofer:2014rta}. LHCb has further observed lepton flavor universality violation (LFUV) in $B\to K\ell^+\ell^-$ decays~\cite{Aaij:2014ora} across the dilepton invariant-mass-squared range $1\,\GeV^2<m_{\ell\ell}^2<6\,\GeV^2$.  Here, the measured branching fraction ratio
\begin{equation}
\label{RK}
R(K)=\frac{\Br[B\to K \mu^+\mu^-]}{\Br[B\to K e^+e^-]}=0.745^{+0.090}_{-0.074}\pm 0.036
\end{equation}
disagrees with the theoretically clean SM prediction $R_{\rm SM}(K)=1.0003 \pm 0.0001$~\cite{Bobeth:2007dw} by $2.6\sigma$. Combining these observables with other $b\to s$ transitions, it is found that NP is preferred over the SM by $4$--$5\sigma$~\cite{Altmannshofer:2015sma,Descotes-Genon:2015uva}. 

Hints for NP of LFUV origin in charged-current $B$ decays were observed for the first time by the BaBar collaboration in $B\to D^{(*)}\tau\nu_\tau$~\cite{Lees:2012xj} in 2012. Recently, these measurements have been confirmed by BELLE~\cite{Huschle:2015rga}, while LHCb has remeasured $B\to D^{*}\tau\nu_\tau$~\cite{Aaij:2015yra}. For the ratio ${R}(X)\equiv \Br[B\to X \tau \nu_\tau]/\Br[B\to X \ell \nu_\ell]$, the current HFAG average~\cite{Amhis:2014hma} of these measurements is
\begin{align}
R(D)_\text{exp}&=\,0.391\pm0.041\pm0.028  \,,\notag\\ 
R(D^*)_\text{exp}&=\,0.322\pm0.018\pm0.012  \,.
\end{align}
Comparing these results to the SM predictions~\cite{Fajfer:2012vx} $R_\text{SM}(D)=0.297\pm0.017$ and $R_\text{SM}(D^*)=0.252\pm0.003$, there is a combined discrepancy of $3.9\sigma$~\cite{Amhis:2014hma}. 

In the Higgs sector, CMS has presented results of a search for the lepton-flavor-violating (LFV) decay mode $h\to\mu\tau$, with a preferred value~\cite{Khachatryan:2015kon}
\begin{equation}
	\Br [h\to\mu\tau] = \left( 0.84_{-0.37}^{+0.39} \right)\% \,,
	\label{h0taumuExp}
\end{equation}
which updates an earlier preliminary result~\cite{CMS:2014hha}. This is consistent with the less precise ATLAS measurement~\cite{Aad:2015gha}, giving a combined significance for NP of $2.6\sigma$, since such a decay is forbidden in the SM. This decay mode is of considerable interest because it hints at LFV in the charged-lepton sector, whereas up to now, LFV has only been observed in the neutrino sector via oscillations. Since the simplest SM extensions that can account for neutrino masses and mixing do not lead to observable $h\to\mu\tau$ rates, the confirmation of this decay would have a significant impact on our understanding of lepton flavor. 

An explanation for $h\to\mu\tau$ can be found by introducing additional scalars~\cite{Campos:2014zaa,Sierra:2014nqa,Heeck:2014qea,Crivellin:2015mga,Dorsner:2015mja,Omura:2015nja,Varzielas:2015joa}, while an explanation for $B\to K^* \mu^+\mu^-$ requires $Z'$ vector bosons~\cite{Descotes-Genon:2013wba,Gauld:2013qba,Buras:2013qja,Gauld:2013qja,Buras:2013dea,Altmannshofer:2014cfa,Glashow:2014iga,Crivellin:2015mga,Crivellin:2015lwa,Niehoff:2015bfa,Sierra:2015fma,Crivellin:2015era,Celis:2015ara,Carmona:2015ena} or leptoquarks~\cite{Gripaios:2014tna,Becirevic:2015asa,Varzielas:2015iva,Alonso:2015sja,Calibbi:2015kma,Bauer:2015knc,Barbieri:2015yvd} to generate current--current interactions like $(\bar{s}\gamma_\alpha P_L b)(\bar{\mu}\gamma^\alpha \mu)$. The tauonic $B$ decays could be explained by charged Higgses~\cite{Crivellin:2012ye,Tanaka:2012nw,Celis:2012dk,Crivellin:2013wna,Crivellin:2015hha}, leptoquarks~\cite{Fajfer:2012jt,Deshpande:2012rr,Sakaki:2013bfa,Alonso:2015sja,Calibbi:2015kma,Bauer:2015knc,Fajfer:2015ycq,Barbieri:2015yvd}, and charged vector bosons~\cite{Greljo:2015mma}. 

In light of these flavor anomalies, we are prompted to consider possible effects of LFUV and LFV in rare kaon decays. One reason to expect correlations between the $B$ meson and kaon sectors concerns the direct $CP$-violating ratio $\epsilon'/\epsilon$.   Recent calculations in the large-$N_c$ limit~\cite{Buras:2015xba,Buras:2015yba} and on the lattice~\cite{Bai:2015nea} suggest that the SM prediction for this quantity falls $2$--$3\sigma$ below the experimental world average $\epsilon^\prime/\epsilon = (16.6\pm2.3)\times 10^{-4}$~\cite{Batley:2002gn,AlaviHarati:2002ye,Worcester:2009qt}.  However, the SM prediction for $\epsilon'/\epsilon$ is sensitive to effects from $\pi\pi$ rescattering in the final state, which are entirely absent in the strict large-$N_c$ limit, while the lattice prediction~\cite{Bai:2015nea} for the $I=0$ phase shift $\delta_0=23.8(4.9)(1.2)^\circ$ is about $3\sigma$ smaller than the value obtained in dispersive treatments~\cite{Colangelo:2001df,GarciaMartin:2011cn,Colangelo:NA62}.  Indeed, combining large-$N_c$ methods with chiral loop corrections can bring the value of $\epsilon'/\epsilon$ in agreement with experiment~\cite{Pallante:2001he,Hambye:2003cy}.

Nevertheless, if the issue of final-state interactions is resolved in the future and the discrepancy persists, then NP contributions due to $Z^\prime$ bosons~\cite{Buras:2015kwd,Buras:2016} or leptoquarks would provide a natural explanation.  In that case, the $B$ meson anomalies and tension in $\epsilon'/\epsilon$ could originate from the same NP, with effects of LFUV and LFV in kaon decays to be expected. In the following, we do not commit ourselves to a specific NP model, but instead focus on the analogous four-fermion operators in the kaon sector which can give the required effect in semileptonic $B$ decays.

For LFUV, the most natural processes to study are $K\to \pi \ell^+\ell^-$ decays since these yield analogous observables to~\eqref{RK}.  However, we also consider the purely leptonic decays $K\to\ell^+\ell^-$ since the electron modes are within experimental reach (unlike $B\to e^+e^-$), and thus these processes are promising probes of NP operators which mediate LFUV. Limits on LFV can be extracted from $K$ decays with $\mu e$ final states. 

The present experimental situation is as follows. For the semileptonic decays, the branching fraction is largest for the charged channels $K^\pm \to \pi^\pm \ell^+\ell^-$, as measured in~\cite{Alliegro:1992pp,Adler:1997zk,Ma:1999uj,Park:2001cv} and studied with high statistics in~\cite{Appel:1999yq,Batley:2009aa,Batley:2011zz}. The PDG averages are~\cite{Agashe:2014kda} 
\begin{align}
  \Br[K^+\to\pi^+e^+e^-] &= (3.00\pm 0.09)\times 10^{-7} \notag \,, \\
  \Br[K^+\to\pi^+\mu^+\mu^-] &= (9.4\pm 0.6)\times 10^{-8}\,,
\end{align}
where the muonic mode includes a scale factor $S=2.6$ of the error due to the conflict with~\cite{Adler:1997zk}.%
  \footnote{Before the remeasurement in~\cite{Ma:1999uj,Park:2001cv,Batley:2011zz}, the result from~\cite{Adler:1997zk} implied a $2\sigma$ tension between the electron and muon decay modes.}
In the neutral-kaon sector the observed decay rates are~\cite{Batley:2003mu,Batley:2004wg} 
\begin{align}
\label{K_S_rates}
  \Br[K_S\to\pi^0e^+e^-]_{m_{ee}>0.165\GeV} &= 3.0^{+1.5}_{-1.2}\times 10^{-9} \,, \notag \\
  \Br[K_S\to\pi^0\mu^+\mu^-] &= 2.9^{+1.5}_{-1.2}\times 10^{-9}\,,
\end{align}
while for the $K_L$ decays only upper limits~\cite{AlaviHarati:2003mr,AlaviHarati:2000hs} are available:
\begin{align}
  \Br[K_L\to\pi^0e^+e^-]<2.8\times 10^{-10} \,, \notag \\
  \Br[K_L\to\pi^0\mu^+\mu^-]<3.8\times 10^{-10} \,.
\end{align}

For the purely leptonic modes, the PDG average for $\Br[K_L\to\mu^+\mu^-]=(6.84\pm 0.11)\times 10^{-6}$ is dominated by the E871 measurement~\cite{Ambrose:2000gj}, and the same experiment reported the sole observation of the electron mode, with branching fraction $\Br[K_L\to e^+e^-]=9^{+6}_{-4}\times 10^{-12}$~\cite{Ambrose:1998cc}.
For later use, these results are conveniently expressed in terms of the ratios 
\begin{equation}
\label{Relldef}
  R_{\ell\ell} = \frac{\Gamma(K_L\to\ell^+\ell^-)}{\Gamma(K_L\to\gamma\gamma)}\,,
\end{equation}
which gives~\cite{Agashe:2014kda}
\begin{align}
\label{Rellexp}
  R_{\mu\mu}^\text{exp} &= (1.25 \pm 0.02) \times 10^{-5}\,,\notag\\
  R_{ee}^\text{exp} &= 1.6^{+1.1}_{-0.7}\times 10^{-8}\,.
\end{align}
We do not consider the related $K_S \to \ell^+\ell^-$ decays, since the SM predictions~\cite{Ecker:1991ru} lie well below the current experimental bounds~\cite{Agashe:2014kda}.  The current limits on the LFV modes are listed in Table~\ref{tab:LFV_exp}.  

\begin{table}[t]
\renewcommand{\arraystretch}{1.3}
\centering
\begin{tabular}{lcr}\toprule
Channel & $\Br$ & Reference \\
\hline
$K^+\to\pi^+\mu^+e^-$ & $<1.3\times 10^{-11}$ & E865, E777~\cite{Sher:2005sp}\\
$K^+\to\pi^+\mu^-e^+$ & $<5.2\times 10^{-10}$ & E865~\cite{Appel:2000tc}\\
$K_L\to\pi^0\mu^\pm e^\mp$ & $<7.6\times 10^{-11}$ & KTeV~\cite{Abouzaid:2007aa}\\
$K_L\to\mu^\pm e^\mp$ & $<4.7\times 10^{-12}$ & E871~\cite{Ambrose:1998us}\\
\botrule
\end{tabular}
\caption{Current limits on branching ratios for LFV decay channels~\cite{Agashe:2014kda}. We do not consider lepton-number-violating modes with $|\Delta L|=2$, whose decay mechanism in general cannot be represented in terms of local operators~\cite{Littenberg:2000fg}.}
\label{tab:LFV_exp}
\end{table}

For the charged $K$ decays, the sensitivity to LFUV and LFV is expected to improve at the high-statistics NA62 experiment~\cite{Hahn:1404985,Moulson:2013oga,Ceccucci:CD2015}, where the nominal number of decays is approximately a factor of  50 larger than that of NA48/2.%
  \footnote{This number refers to the best-case scenario where no downscaling of the rare decay trigger chains is imposed.  For modes like $K^+ \to \pi^+ e^+ e^-$, downscaling factors as large as 10 are foreseen~\cite{Evgueni}, so that the statistics increase may be reduced to a factor of 5.}
 For example, the projected limit for $\Br[K^+\to\pi^+\mu^+ e^-]$ becomes $0.7\times 10^{-12}$.   For $K_L$ decays, the KOTO experiment at J-PARC~\cite{Komatsubara:2012pn,Nanjo:2015cra} has good prospects of reaching SM sensitivity for $K_L \to \pi^0\nu\bar\nu$. In principle, the increased reach might be sufficient to probe the $K_L$ modes involving charged lepton pairs, but the detection of these final states would require a different search strategy to the one employed for $K_L \to \pi^0\nu\bar\nu$. Finally, although we restrict our focus to the neutral-current sector, there is also renewed interest in charged-current processes at the J-PARC E36 experiment, which is searching for signs of LFUV in $K^+ \to \ell^+\nu_\ell$~\cite{Shimizu:2015xzf}.

On the theory side, all $K$ decays have been studied thoroughly in the context of chiral $SU(3)_L \times SU(3)_R$ perturbation theory ($\chi$PT$_3$)~\cite{Ecker:1987qi,Ecker:1987hd,Donoghue:1994yt,D'Ambrosio:1998yj,GomezDumm:1998gw,Knecht:1999gb,Buchalla:2003sj,Isidori:2003ts,Isidori:2004rb,Kubis:2010mp,Mertens:2011ts,Ananthanarayan:2012hu}, with the present status reviewed in \cite{Cirigliano:2011ny}. The general picture that arises is the presence of long-distance physics, parametrized in terms of low-energy constants (LECs) in the effective weak Lagrangian. The values of these LECs are poorly known in most cases, and this limits the predictive power of $\chi$PT$_3$ in the weak sector. However, information on short-distance physics can be extracted by considering decay spectra as well as interrelations among different decay modes. Furthermore, LFV decay channels are typically less affected by hadronic uncertainties, and have been used in the past to extract limits on the NP scale~\cite{Littenberg:1993qv}. Recently, the prospects of calculating the long-distance contributions on the lattice have been discussed~\cite{Christ:2015aha}, although it will take several years before high precision is reached.

This article is organized as follows.  In Sec.~\ref{sec:formalism} we establish our conventions and the general formalism necessary to study leptonic and semileptonic $K$ decays.  LFUV in $K\to\pi\ell^+\ell^-$ decays is analyzed in Sec.~\ref{sec:Kpill}, where the assumption of minimal flavor violation (MFV)~\cite{Chivukula:1987fw,Hall:1990ac,Buras:2000dm,D'Ambrosio:2002ex,Bobeth:2005ck} is used 
to relate experimental limits in $K$ and $B$ decays. LFUV in the purely leptonic modes is discussed in Sec.~\ref{sec:KLll}, while the LFV decays are discussed in Sec.~\ref{sec:LFV}.  We conclude in Sec.~\ref{sec:conclusion}.

\section{Formalism}
\label{sec:formalism}

\begin{figure}[t]
  \centering\includegraphics[width=\linewidth]{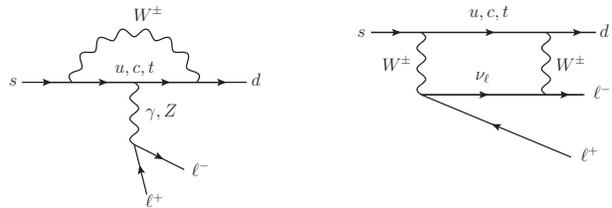}
  \caption{One-loop graphs which give a short-distance contribution to $K\to\pi\ell^+\ell^-$.}
  \label{fig:penguin}
\end{figure}

We follow the notation and conventions from~\cite{Cirigliano:2011ny}. To leading order in $m_W^{-2}$ and inverse heavy quark masses, the $|\Delta S|=1$ interactions are defined by the effective Lagrangian
\begin{equation}
\label{Lagr_K}
  {\cal L}_\text{eff}^{|\Delta S|=1} = 
	- \frac{G_F}{\sqrt{2}} V_{ud}V_{us}^* \sum_{i=1}^{13} C_i(\mu) Q_i(\mu) + \text{h.c.} \,,
\end{equation}
where $\{Q_i\}$ is a set of local composite operators with Wilson coefficients $C_i$. For the rare $K$ decays under consideration, the relevant energy scale is $\mu \ll m_{t,c,b}$, so we only need the four-quark operators $Q_{1\text{--}6}$
\begin{align}
\label{eq:Q1-6}
Q_1  & =   \left[\bar s^\alpha \gamma^\mu (1 - \gamma_5) u^\beta\right] \left[\bar u^\beta \gamma_\mu (1 - \gamma_5) d^\alpha\right], \notag\\
Q_2 & = 
\left[\bar s\gamma^\mu (1 - \gamma_5) u\right] \ \left[\bar u \gamma_\mu (1 - \gamma_5) d\right]\,, \notag\\
Q_3 & =   \left[\bar s \gamma^\mu (1 - \gamma_5) d\right]  \sum_{q=u,d,s} \left[\bar q \gamma_\mu (1 - \gamma_5) q \right]\,, \notag\\
Q_4 & =   \left[\bar s^\alpha \gamma^\mu (1 - \gamma_5) d^\beta\right]
\sum_{q=u,d,s} \left[\bar q^\beta \gamma_\mu (1 - \gamma_5) q^\alpha\right]\,, \notag\\
Q_5 & =  \left[ \bar s \gamma^\mu (1 - \gamma_5) d\right]
\sum_{q=u,d,s} \, \left[\bar q \gamma_\mu (1 + \gamma_5) q\right]\,,\notag\\
Q_6 & =  \left[\bar s^\alpha \gamma^\mu (1 - \gamma_5) d^\beta\right]
 \sum_{q=u,d,s} \left[\bar q^\beta \gamma_\mu (1 + \gamma_5) q^\alpha\right]\,,
\end{align}
as well as the Gilman--Wise operators~\cite{Gilman:1979ud,Inami:1980fz,Dib:1988js,Dib:1988md,Flynn:1988ve}
\begin{align}\label{eq:Q11-12}
Q_{11} \equiv  Q_{7V} &=   \left[\bar s \gamma^\mu (1 - \gamma_5) d \right]
\sum_{\ell=e,\mu} \left[\bar{\ell} \gamma_\mu  \ell \right], \notag\\
Q_{12} \equiv Q_{7A} & = \left[\bar s \gamma^\mu (1 - \gamma_5) d \right]
\sum_{\ell=e,\mu}  \left[\bar{\ell} \gamma_\mu \gamma_5 \ell \right]\,. 
\end{align}
We use $\alpha$, $\beta$ to denote color indices; otherwise the Dirac bilinears $\bar{f}\Gamma f$ are understood to be color singlets. For the Wilson coefficients we adopt the standard decomposition
\begin{equation}
\label{C_i_dec}
  C_i(\mu) = z_i(\mu) + \tau y_i(\mu)\,, \qquad \tau = - \frac{V_{td}V_{ts}^*}{V_{ud}V_{us}^*} \,,
\end{equation}
which arises from first decoupling $t,W,Z$ simultaneously at $\mu=m_W$, followed by successively integrating out the $b$ and $c$ quarks in the evolution from $\mu=m_W$ to $\mu \lesssim m_c$~\cite{Buras:1994qa}.
At zeroth order in the strong interactions and to $O(g^2)$ in the weak interactions, $C_2$ is the only nonvanishing Wilson coefficient.  At $O(e^2)$, the $\gamma,Z$-penguin and $W$-box graphs in Fig.~\ref{fig:penguin} generate nonzero coefficients for $Q_{7V}$ and $Q_{7A}$~\cite{Inami:1980fz}, while $O(g_s^2)$ corrections generate nonzero contributions to $C_{1\text{--}6}$.

Note that we have assumed right-handed quark currents are absent, as in the SM.  This is because symmetry-based solutions to the anomalies in semileptonic $B$ decays include 1) a left-handed $\bar sb$ current and a vectorial muon current, and 2) a left-handed $\bar sb$ current and a left-handed muon current.  This pattern suggests NP effects involving similar operators in kaon decays.

The calculation of $K \to \pi\ell\ell'$ and $K\to\ell\ell'$ amplitudes involves hadronic matrix elements such as $\langle \gamma^*\pi | {\cal L}_\mathrm{eff} | K\rangle$, whose determination requires nonperturbative methods. These matrix elements can be systematically analyzed in $\chi$PT$_3$, where amplitudes are expanded in powers of $O(\mk)$ momenta $p$ and quark masses $m_{u,d,s}=O(\mk^2)$ (with $m_{u,d}/m_s$ held fixed). For $|\Delta S|=1$ transitions, the content of these calculations is summarized by an effective weak Lagrangian, constrained by the requirements of approximate chiral $SU(3)_L\times SU(3)_R$ symmetry and a discrete $CPS$ symmetry~\cite{Bernard:1985wf}, which interchanges the $s$ and $d$ quarks.  The result is a set of effective weak operators which transform in the same way as ${\cal L}_\mathrm{eff}$, i.e.\ in the $(8_L,1_R)$ and $(27_L,1_R)$ representations of the chiral group.  

Empirically, it is observed that $\Delta I =1/2$ transitions dominate nonleptonic processes, which in $\chi$PT$_3$ corresponds to dominance by octet operators.  It is not clear how this fact should be accounted for, although explanations based on large-$N_c$~\cite{Bardeen:1986vp,Bardeen:1986vz,Buras:2014maa,Pich:1995qp} or an infrared fixed point in the three-flavor strong coupling~\cite{Crewther:2012wd,Crewther:2013vea} have been proposed.\footnote{A direct determination of the $K\to \pi\pi$ amplitudes is not sufficient to explain the $\Delta I =1/2$ rule, since one cannot disentangle contact terms from effects due to final-state rescattering.  Recently, a proposal~\cite{Crewther:2015dpa} to separate these two contributions has been presented, based on a lattice measurement of $K \to \pi$ on-shell.}

In the context of potential NP contributions to $C_{7V}$ and $C_{7A}$, one needs the chiral realization of the octet quark operator.  At lowest order in $\chi$PT$_3$, this is obtained by projecting the usual $SU(3)_L$ chiral current $\sim U\partial_\mu U^\dagger$ onto the $\Delta S=-1$ sector~\cite{Ecker:1987hd}:%
  \footnote{Note that the relation (\ref{bosonize}) only relies on chiral symmetry. Large-$N_c$ arguments~\cite{Ecker:1987hd,Bruno:1992za} are needed only if a relation between the Gilman--Wise operators and corresponding LECs of the effective weak Lagrangian is sought.}
\begin{equation}
  \bar{s}\gamma^\mu(1-\gamma_5)d \leftrightarrow i F_0^2 (U\partial^\mu U^\dagger)_{23}\,.
  \label{bosonize}
\end{equation}
Here, $U=U(\pi,K,\eta)$ is a chiral $SU(3)$ field, and $F_0$ is the meson decay constant in the chiral limit, whose value can be determined from either the pion or the kaon channel. (Numerically, we use $F_\pi=92.2\MeV$ and $F_K/F_\pi=1.22$~\cite{Agashe:2014kda}.)

For later convenience we also quote the analogous conventions for $B$ decays~\cite{Descotes-Genon:2015uva}
\begin{equation}
\label{H_B}
  {\cal H}_\text{eff}^{|\Delta B|=1} = - \frac{4G_F}{\sqrt{2}} V_{tb}V_{ts}^* \sum_{i} C_i^B(\mu) Q_i^B(\mu)+ \text{h.c.}\,,
\end{equation}
where
\begin{align}
\label{Qi_B}
 Q_{9}^B &=   \frac{e^2}{32\pi^2}\left[\bar s \gamma^\mu (1 - \gamma_5) b \right]
\sum_{\ell=e,\mu} \left[\bar{\ell} \gamma_\mu  \ell \right]\,, \notag\\
Q_{10}^B & = \frac{e^2}{32\pi^2}\left[\bar s \gamma^\mu (1 - \gamma_5) b \right]
\sum_{\ell=e,\mu}  \left[\bar{\ell} \gamma_\mu \gamma_5 \ell \right]\,.
\end{align}

\section{LFUV in semileptonic $\boldsymbol{K}$ decays}
\label{sec:Kpill}

\subsection{$\boldsymbol{K^\pm \to \pi^\pm \ell^+\ell^-}$ decays}
\label{sec:charged}

At low energies, the dominant $CP$-conserving contribution to 
\begin{equation}
  K^+(k) \to \pi^+(p)\ell^+(p_+)\ell^-(p_-) \,, \qquad \ell = \mu \mbox{ or } e\,,
\end{equation}
is known~\cite{Ecker:1987qi} to arise from single virtual-photon exchange%
  \footnote{In $K_L\to \pi^0\ell^+\ell^-$ decays, this contribution is $CP$-violating; see Sec.\ \ref{sec:KL}.}
\begin{equation}
  K^+(k) \to \pi^+(p) \gamma^*(q,\lambda)\,, \quad q = k-p\,, \quad q^2 = m_{\ell\ell}^2\,,
  \label{photon exchange}
\end{equation}
where $\lambda$ denotes the polarization of the photon. Barring the $\Delta I=1/2$ rule, there is no rigorous theoretical argument why~\eqref{photon exchange} should dominate; after all, there are short-distance contributions from $Z$-penguin and $W$-box diagrams (Fig.~\ref{fig:penguin}).  Moreover, it is not possible to make a clean theoretical prediction for the $\gamma$-penguin contribution $C^\gamma_{7V}$ associated with $C_{7V}$.  As noted in~\cite{Dib:1988js,Dib:1988md}, the QCD corrections to $C^\gamma_{7V}$ for $t$ and $c$ quarks are large and change both the magnitude and sign of the Wilson coefficient. Nevertheless, a rough estimate of the rate $K\to\pi\ell^+\ell^-$ due to an amplitude $\sim C_{7V}$ gives a result far too small to explain the data. It is on this basis that short-distance contributions from $Q_{7V}$ (as well as $Q_{7A}$) are typically neglected in calculations of the branching ratios and spectra.

The photon contribution~\eqref{photon exchange} gives rise to the amplitude 
\begin{equation}
  A_\text{V}^{K^+\to \pi^+\ell^+\ell^-} = -\frac{G_F\alpha}{4\pi} V_+(z)\bar{u}_\ell(p_-)(\slashed{k}+\slashed{p}) v_\ell(p_+)\,,
\end{equation}
where $V_+(z)$ is the vector form factor and $z=q^2/\mk^2$ is the momentum transfer.  In the physical region
$4r_\ell^2 \leq z \leq (1-r_\pi)^2$,  $r_i = m_i / \mk$,
the differential decay rate is 
\begin{equation}
  \frac{\diff\Gamma}{\diff z} = \frac{G_F^2\alpha^2\mk^5}{12\pi(4\pi)^4} \bar{\lambda}^{3/2} \sqrt{1-4\frac{r_\ell^2}{z}} \bigg(1+2\frac{r_\ell^2}{z} \bigg) |V_+(z)|^2\,,
  \label{dGdz}
\end{equation}
where $\bar{\lambda} = \lambda(1,z,r_\pi^2)$ and  $\lambda(a,b,c) = a^2 + b^2 + c^2 - 2(ab+bc+ac)$.

The requirements of chiral symmetry and gauge invariance imply that $V_+(z)$ vanishes at $O(p^2)$ in $\chi$PT$_3$~\cite{Ecker:1987qi}, so the lowest-order contribution occurs at $O(p^4)$. Beyond $O(p^4)$, $\pi\pi$ rescattering in the nonleptonic decay $K\to\pi\pi\pi$ needs to be taken into account as well~\cite{D'Ambrosio:1998yj}. Given the limited information on most of the LECs, it is convenient to adopt a general representation~\cite{D'Ambrosio:1998yj} of the form factor
\begin{equation}
  V_+(z) = a_+ + b_+z + V_+^{\pi\pi}(z)\,,
  \label{V_dec}
\end{equation}
which is valid at $O(p^6)$.  Here $a_+$ and $b_+$ parametrize the polynomial part, while the rescattering contribution $V_+^{\pi\pi}$ can be determined from fits to $K\to\pi\pi$ and $K\to\pi\pi\pi$ data~\cite{Kambor:1991ah,Bijnens:2002vr}. In general, $V_+$ receives contributions from both the octet and 27-plet parts of ${\cal L}_\text{eff}$~\cite{Ananthanarayan:2012hu}, although the $\Delta I = 1/2$ rule implies octet dominance, and thus the latter contributions are generally suppressed. 

\begin{table}[t]
\renewcommand{\arraystretch}{1.3}
\centering
\begin{tabular}{cccr}\toprule
Channel & $a_+$ & $b_+$ & Reference \\
\hline
$ee$ & $-0.587\pm 0.010$ & $-0.655\pm 0.044$ & E865~\cite{Appel:1999yq}\\
$ee$ & $-0.578\pm 0.016$ & $-0.779\pm 0.066$ & NA48/2~\cite{Batley:2009aa}\\
$\mu\mu$ & $-0.575\pm 0.039$ & $-0.813\pm 0.145$ & NA48/2~\cite{Batley:2011zz}\\\botrule
\end{tabular}
\caption{Coefficients in the vector form factor~\eqref{V_dec}.}
\label{tab:a+b+}
\end{table}

The representation~\eqref{V_dec} was used as a fit function in all available high-statistic experiments with the results given in Table~\ref{tab:a+b+}. If LFU holds, the coefficients have to be equal for the electron and muon channels, which within errors is indeed the case.%
  \footnote{Although note a small $1.6\sigma$ tension in the $b_+$ coefficient between the two electron experiments.
  We define LFU in the usual sense, i.e.\ excluding the Yukawa interactions in the SM (otherwise the different lepton masses would break LFU trivially).} 
Any discrepancy can then be attributed to NP, and thus the corresponding effect would be necessarily short-distance. It follows that the $O(p^2)$ chiral realization~\eqref{bosonize} of the $Q_{7V}$ operator converts the allowed range in $a_+^\text{NP}$ into a corresponding range in the Wilson coefficients~\cite{Ecker:1987hd} 
\begin{equation}
 a_+^\text{NP}=\frac{2\pi\sqrt{2}}{\alpha}V_{ud}V_{us}^*C_{7V}^\text{NP}\,,
 \label{aNP}
\end{equation}
and thus the difference between the two channels is
\begin{equation}
\label{limit_Kp}
 C_{7V}^{\mu\mu}-C_{7V}^{ee}=\alpha\frac{a_+^{\mu\mu}-a_+^{ee}}{2\pi\sqrt{2}V_{ud}V_{us}^*}\,.
\end{equation}
If we assume MFV (to be understood in its simplest form, i.e.\ as the first order in the expansion in~\cite{D'Ambrosio:2002ex}), this translates into a constraint on the NP contribution to $C_9^B$:
\begin{equation}
\label{C_charged}
 C_{9}^{B,\mu\mu}-C_{9}^{B,ee}=-\frac{a_+^{\mu\mu}-a_+^{ee}}{\sqrt{2}\lambda_t}\approx -19\pm 79\,,
\end{equation}
where we have averaged over the two electron experiments, defined $\lambda_t=V_{ts}^*V_{td}$, and used PDG global-fit values for the CKM matrix elements~\cite{Agashe:2014kda}.%
  \footnote{In the estimate~\eqref{C_charged} we did not include effects due to renormalization group running between the scales of $B$-physics and $\chi$PT$_3$. However, the semileptonic operators involve a vector or axial-vector current, so they are not renormalized (at the one-loop level). There is only a mixing of four-quark operators into the semileptonic operators, which is LFU conserving.} 
In particular, we may use the modulus of $\lambda_t$ in \eqref{C_charged} since MFV implies that the respective phases coincide with the SM, so that $C_{7V}/C_{7V}^\text{SM}=C_{9}^B/C_{9}^{B,\text{SM}}$ and $C_{7A}/C_{7A}^\text{SM}=C_{10}^B/C_{10}^{B,\text{SM}}$ (the remaining factors are simply due to the different normalizations of the effective Hamiltonians). 

Evidently, the determination of $a_+^{\mu\mu}-a_+^{ee}$ would need to be improved by at least an order of magnitude to probe the parameter space relevant for the $B$ anomalies~\cite{Descotes-Genon:2015uva}, whose explanation involves Wilson coefficients $C_{9,10}^B=O(1)$. Progress in this direction can be anticipated at NA62, especially for the experimentally cleaner dimuon mode which currently has the larger uncertainty. It should be stressed that if NP does not satisfy MFV, the relative size of NP contributions to the Wilson coefficients is not fixed. In this case it is possible that the relative NP effects in the kaon sector are larger than in the $B$ meson decays because the short-distance SM contribution is CKM suppressed in the former.

\begin{figure}[t]
  \centering\includegraphics[width=\linewidth]{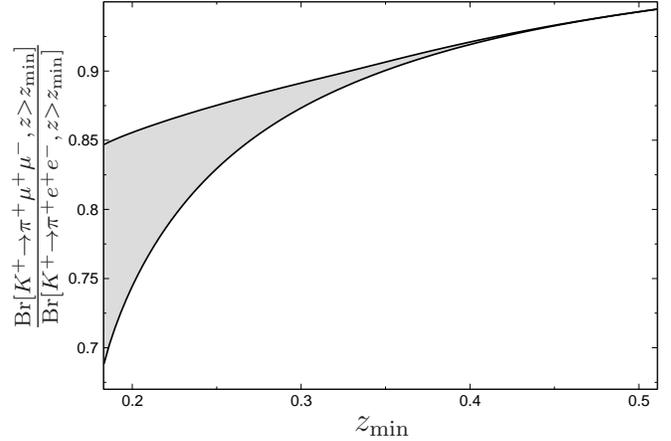}
  \caption{Ratio of muon and electron branching fractions for $4r_\mu^2 \leq z_\text{min} \leq (1-r_\pi)^2$ and $a_+,b_+\in[-1,0]$.}
  \label{fig:ratio}
\end{figure}

An alternative analysis strategy, often applied in $B$ decays, to minimize sensitivity to hadronic form factors~\cite{Bobeth:2007dw} relies on the ratio of branching fractions 
\begin{equation}
  \frac{\Br[K^+\to\pi^+\mu^+\mu^-,z>z_\text{min}]}{\Br[K^+\to\pi^+e^+e^-,z>z_\text{min}]}\,,
\end{equation}
where $z_\mathrm{min}$ is a cutoff on the spectrum.  
While the impact of the muon mass is negligible in the $B$-physics case, this is not true for kaons and a lower $z_\text{min}$ must be applied to reduce the theory uncertainties. Indeed, as shown in Fig.~\ref{fig:ratio}, for given ranges in $a_+$ and $b_+$ the uncertainty in the ratio decreases quickly with increasing $z_\text{min}$. However, in practice the determination of the ranges in the coefficients still requires a fit to the spectrum, so that all information on LFU can equivalently be extracted from this fit. 

It has been observed in~\cite{D'Ambrosio:1998yj} that the long-distance contributions could also be eliminated in the $CP$-violating charge asymmetry
\begin{equation}
A_{CP}^{\ell\ell}=\frac{\Gamma[K^+\to\pi^+\ell^+\ell^-]-\Gamma[K^-\to\pi^-\ell^+\ell^-]}{\Gamma[K^+\to\pi^+\ell^+\ell^-]+\Gamma[K^-\to\pi^-\ell^+\ell^-]}\,,
\label{asymm}
\end{equation}
which in the SM is determined by $\Im \lambda_t$. Taking $\Im\lambda_t=1.35\times 10^{-4}$, the resulting SM prediction for \eqref{asymm} is $\sim 10^{-5}$~\cite{D'Ambrosio:1998yj}.  This is to be compared with the most stringent experimental constraints $A_{CP}^{ee}=(-2.2\pm 1.6)\times 10^{-2}$~\cite{Batley:2009aa}, and $A_{CP}^{\mu\mu}=(1.1\pm 2.3)\times 10^{-2}$~\cite{Batley:2011zz}, so we conclude that reaching SM sensitivity would require an improvement by 3 orders of magnitude.

In principle, there are additional axial-vector contributions to $K^+\to\pi^+\ell^+\ell^-$, e.g.\ due to $Z$ exchange (Fig.\ \ref{fig:penguin}) or NP mediators like $Z'$ bosons or leptoquarks.  This contribution generates an amplitude of the form
\begin{equation}
  A_A^{K^+\to\pi^+\ell^+\ell^-} = -\frac{G_F\alpha}{4\pi}A_+(z) \bar u_\ell (\slashed{k}+\slashed{p})\gamma_5 v_\ell\,,
\end{equation}
where by analogy with \eqref{V_dec}, we take the lowest order decomposition $A_+(z) = d_+$ for the axial-vector form factor.  Redoing the fit in terms of $A = A_V + A_A$,
\begin{align}
  \frac{d\Gamma}{dz} &= \frac{G_F^2\alpha^2\mk^5}{12\pi(4\pi)^4} \sqrt{\bar{\lambda}} \sqrt{1-4\frac{r_\ell^2}{z}} \Bigg\{6r_\ell^2\big(2+2r_\pi^2-z\big)|A_+(z)|^2\notag\\
  &+\bar \lambda\bigg(1+2\frac{r_\ell^2}{z} \bigg) \Big(|V_+(z)|^2+|A_+(z)|^2\Big)\Bigg\}\,,
\end{align}
gives $d_+^{ee} = 0.00\pm 0.47$ and $d_+^{\mu\mu}=0.00\pm 0.13$, which in turn yields the very weak bound $|C_{10}^{B,\mu\mu} - C_{10}^{B,ee}| \lesssim 1000$. 

One critical factor in improving the accuracy of~\eqref{C_charged} concerns radiative corrections, which
in~\cite{Appel:1999yq,Batley:2009aa,Batley:2011zz} were performed according to the leading Coulomb factor~\cite{Lautrup:1971ew,Isidori:2007zt}. More recently, these corrections have been addressed in full detail in a $\chi$PT$_3$ calculation assuming a linear form factor~\cite{Kubis:2010mp}, in particular demonstrating that the corrections to the decay spectrum can still be expressed in a factorized form. These results should be valuable in view of the expected increase in statistics in the NA62 experiment.

While the extraction of short-distance physics from $K^+\to\pi^+\ell^+\ell^-$ decays themselves is difficult, a more precise measurement of its decay spectrum would have indirect implications for $K_{S,L}\to\pi^0 \ell^+\ell^-$: the numerical value of $b_+$ is larger than expected from dimensional counting or vector meson dominance (VMD), where the latter predicts $b_+/a_+=1/r_V^2=\mk^2/M_\rho^2\simeq 0.4$. With increased statistics one might become sensitive to a quadratic term $\sim c_+ z^2$ in the  expansion of the form factor (\ref{V_dec}), and thereby test the hypothesis that VMD ought to be a decent description of $V_+$ once a non-VMD portion in $a_+$ related to sizable pion-loop contributions in this channel is subtracted~\cite{D'Ambrosio:1998yj,Buchalla:2003sj}. Arguments along these lines are used to justify VMD assumptions in $K_S\to\pi^0 \ell^+\ell^-$, and, thereby, help fix the relative sign of the interference term between direct and indirect $CP$-violating contributions in $K_L\to\pi^0 \ell^+\ell^-$~\cite{Buchalla:2003sj}. 

\subsection{$\boldsymbol{K_S \to \pi^0 \ell^+\ell^-}$ decays}
\label{sec:KS}

The expression for the $K_S \to \pi^0 \ell^+\ell^-$ spectrum is very similar to~\eqref{dGdz}, with neutral particle masses in the phase space expression and parameters $a_+$, $b_+$ replaced by $a_S$, $b_S$ in the form factor. 
Since the nonleptonic mode $K_S \to \pi\pi$ dominates the total $K_S$ width, the branching fraction for $K_S \to \pi^0 \ell^+\ell^-$ is smaller than for the charged decay, and it is even more difficult to directly extract information on short-distance physics. However, a measurement of the spectrum would enable an explicit test of the VMD assumption for $b_S/a_S = 1/r_V^2$, which is expected to work better than in the charged channel due to the lesser role of pion loops. Use of the VMD assumption and the decay rates~\eqref{K_S_rates} implies that $a_S$ is only known with large uncertainties~\cite{Cirigliano:2011ny}:
\begin{equation}
|a_S^{ee}|=1.06^{+0.26}_{-0.21}\,,\qquad
|a_S^{\mu\mu}|=1.54^{+0.40}_{-0.32}\,.
\end{equation}
As we discuss in the next subsection, any additional information on $K_S \to \pi^0 \ell^+\ell^-$ would sharpen the prediction of the indirect $CP$-violating contribution to $K_L \to \pi^0 \ell^+\ell^-$.

\subsection{$\boldsymbol{K_L \to \pi^0 \ell^+\ell^-}$ decays}
\label{sec:KL}

The process $K_L \to \pi^0 \ell^+\ell^-$ is driven by three different decay mechanisms: a direct $CP$-violating\footnote{$K_L\to\pi^0(\ell^+\ell^-)_{J=1}$ with a vector or axial-vector lepton pair is $CP$-violating~\cite{Cirigliano:2011ny}.} amplitude of short-distance origin from $Q_{7V}$ and $Q_{7A}$, an indirect $CP$-violating transition due to $K^0$--$\bar K^0$ oscillations, and a $CP$-conserving contribution originating from $K_L\to\pi^0\gamma\gamma$ and subsequent $\gamma\gamma\to\ell^+\ell^-$ rescattering (with $J=0,2,\ldots$ two-photon states).
The corresponding form of the decay spectrum as well as the consequences for extracting short-distance physics have been investigated in detail in~\cite{Ecker:1987hd,Buras:1994qa,Donoghue:1994yt,Buchalla:2003sj,Isidori:2004rb,Mescia:2006jd}; here we review the salient features. First, the decay spectrum for the $CP$-violating part takes the form
\begin{align}
 \frac{\diff\Gamma}{\diff z}\bigg|_{\text{CPV}}&=\frac{G_F^2\alpha^2\mk^5}{12\pi(4\pi)^4} \sqrt{\bar{\lambda}} \sqrt{1-4\frac{r_\ell^2}{z}}\Bigg\{
 \frac{3}{2}r_\ell^2 z|P_0(z)|^2\notag\\
 &+\bar{\lambda}\bigg(1+2\frac{r_\ell^2}{z} \bigg) |V_0(z)|^2\notag\\
 &+\bigg[\bar{\lambda}\bigg(1+2\frac{r_\ell^2}{z} \bigg)+6r_\ell^2\big(2+2r_\pi^2-z\big)\bigg]|A_0(z)|^2\notag\\
 &+6r_\ell^2\big(1-r_\pi^2\big)\Re\big[A_0(z)^*P_0(z)\big]\Bigg\}\,,
\end{align}
which in the limit of a purely vector interaction reduces to the neutral-channel analog of~\eqref{dGdz}. The vector, axial-vector, and pseudoscalar amplitudes are defined as
\begin{align}
  A_\text{V}^{K_L\to \pi^0\ell^+\ell^-} &= -\frac{G_F\alpha}{4\pi} V_0(z)\bar{u}_\ell(p_-)(\slashed{k}+\slashed{p}) v_\ell(p_+)\,,\notag\\
  A_\text{A}^{K_L\to \pi^0\ell^+\ell^-} &= -\frac{G_F\alpha}{4\pi} A_0(z)\bar{u}_\ell(p_-)(\slashed{k}+\slashed{p})\gamma_5 v_\ell(p_+)\,,\notag\\
  A_\text{P}^{K_L\to \pi^0\ell^+\ell^-} &= -\frac{G_F\alpha}{4\pi} P_0(z)m_\ell\bar{u}_\ell(p_-)\gamma_5 v_\ell(p_+)\,.
\end{align}
Indirect $CP$ violation leads to a vector amplitude of the form
\begin{equation}
 V_0^\text{indirect}(z)=\eps(a_S+b_S z)\sim \eps a_S\Big(1+\frac{z}{r_V^2}\Big)\,,
\end{equation}
where $\eps\sim e^{i\pi/4}|\eps|$ parametrizes $K^0$--$\bar K^0$ mixing, the $\pi\pi$ rescattering corrections have been neglected, and the second relation follows if VMD is assumed for the polynomial part.

Short-distance physics only affects the direct $CP$-violating contributions
\begin{align}
 V_0^\text{direct}(z)&=i\frac{2\pi\sqrt{2}\,y_{7V}}{\alpha}f_+^{K\pi}(z)\Im\lambda_t\,,\notag\\
 A_0^\text{direct}(z)&=i\frac{2\pi\sqrt{2}\,y_{7A}}{\alpha}f_+^{K\pi}(z)\Im\lambda_t\,,\notag\\
 P_0^\text{direct}(z)&=-i\frac{4\pi\sqrt{2}\,y_{7A}}{\alpha}f_-^{K\pi}(z)\Im\lambda_t\,,
\end{align}
with Wilson coefficients as defined in~\eqref{C_i_dec} and $K_{\ell 3}$ form factors $f_{\pm}^{K\pi}(z)$.
Using the form-factor normalization $f_+(0)$ from~\cite{Aoki:2013ldr}, the slopes from~\cite{Antonelli:2008jg}, $y_{7V,7A}$ from~\cite{Buchalla:2003sj}, and PDG input for the remaining parameters, we obtain for the decay rates
\begin{align}
 \label{CPV}
 &\Br[K_L\to\pi^0e^+e^-]\big|_\text{CPV}\notag\\
 &=10^{-12}\bigg[14.8|a_S|^2\pm 6.2|a_S|\bigg(\frac{\Im\lambda_t}{10^{-4}}\bigg)+2.5\bigg(\frac{\Im\lambda_t}{10^{-4}}\bigg)^2\bigg]\,,\notag\\
 &\Br[K_L\to\pi^0\mu^+\mu^-]\big|_\text{CPV}\notag\\
 &=10^{-12}\bigg[3.5|a_S|^2\pm 1.5|a_S|\bigg(\frac{\Im\lambda_t}{10^{-4}}\bigg)+1.1\bigg(\frac{\Im\lambda_t}{10^{-4}}\bigg)^2\bigg]\,.
\end{align}
More precise information on $K_S\to\pi^0 \ell^+\ell^-$ would be highly beneficial for several reasons all related to the indirect $CP$-violating part of~\eqref{CPV}: its derivation relies on the VMD assumption for $b_S$.  As it stands, the dominant uncertainty resides in $a_S$ and the arguments put forward in~\cite{Buchalla:2003sj} in favor of a positive sign of the interference term rely on the separation of VMD and non-VMD contributions to the polynomial coefficients, assumptions that could be tested with more precise data on $K_S\to\pi^0 \ell^+\ell^-$ (and also $K^\pm\to\pi^\pm \ell^+\ell^-$). The $CP$-conserving contribution to the muon channel has been estimated to be~\cite{Isidori:2004rb}
\begin{equation}
\label{CP_cons}
 \Br[K_L\to\pi^0\mu^+\mu^-]\big|_\text{CPC}=(5.2\pm 1.6)\times 10^{-12},
\end{equation}
which is of the same order of magnitude as the $CP$-violating part. The $CP$-conserving electron decay channel is further suppressed~\cite{Ecker:1987hd,Isidori:2004rb,Cirigliano:2011ny}. 

Comparing~\eqref{Lagr_K}--\eqref{Qi_B}, MFV suggests the identification $y_{7V,7A}\sim C_{9,10}^B\alpha/2\pi$, so that a NP contribution to $C_{9,10}^B=O(1)$ would imply $y_{7V,7A}=O(10^{-3})$, about a factor of $5$ less than the SM values of $y_{7V,7A}$. For $a_S=1$, the $CP$-violating branching fractions become 
\begin{align}
  \Br[K_L\to\pi^0e^+e^-]|_\text{CPV} &= 2.8\times 10^{-11}\,, \notag \\
  \Br[K_L\to\pi^0\mu^+\mu^-]|_\text{CPV} &= 7.4\times 10^{-12}\,.
\end{align}
Starting from this benchmark point, shifts in $y_{7V}$ by $\pm 10^{-3}$ with $y_{7A}$ held fixed (and vice versa) produce effects in the windows $[2.5,3.0]\times 10^{-11}$ and $[6.9,8.0]\times 10^{-12}$, respectively, which in the case of the muon channel is even less than the uncertainty in the $CP$-conserving contribution~\eqref{CP_cons}.
If NP were to obey MFV, a test of the $B$-physics anomalies in $K_L\to\pi^0\ell^+\ell^-$ therefore appears very challenging.

\section{$\boldsymbol{K_L \to \ell^+\ell^-}$ Decays}
\label{sec:KLll}

In Sec.~\ref{sec:charged} we saw that the $K\to \pi\ell^+\ell^-$ decays provided a probe of LFUV in NP scenarios involving vector-current interactions. Here we examine the complementary role provided by $K_L\to\ell^+\ell^-$ in constraining NP effects due to axial-vector interactions.%
  \footnote{In general, scalar operators of the form $\sim \bar s d \bar \ell \ell$ and $\sim \bar s d \bar \ell \gamma_5 \ell$ (and their pseudoscalar counterparts) could also generate new sources of LFUV.  However, since our analysis is motivated by the anomalies in the $B$ meson sector, which can be explained by (axial-)vector currents but not (pseudo)scalar ones, we do not consider (pseudo)scalar currents here.}
In these decays, there are both long- and short-distance contributions, with the former dominated by $K_L\to\gamma^*\gamma^*\to\ell^+\ell^-$.  As a result, it is convenient to normalize $\Gamma(K_L\to\ell^+\ell^-)$ to the $K_L\to\gamma\gamma$ rate~\eqref{Relldef}, which can be expressed as
\begin{equation}
  R_{\ell\ell}= 2\beta_\ell \bigg(\frac{\alpha}{\pi}r_\ell\bigg)^2 
  \big( |F_{\ell,\text{abs}}|^2 + |F_{\ell,\text{disp}}|^2 \big)\,,
  \label{RL}
\end{equation}
where $\beta_\ell = \sqrt{1-4r_\ell^2}$ and the absorptive and dispersive components are~\cite{Martin:1970ai,D'Ambrosio:1997jp,GomezDumm:1998gw,Knecht:1999gb,Isidori:2003ts}
\begin{align}
  F_{\ell,\text{abs}} &= \frac{\pi}{2\beta_\ell}\log \bigg(\frac{1-\beta_\ell}{1+\beta_\ell}\bigg)\,, \notag \\
  F_{\ell,\text{disp}} &= \frac{1}{4\beta_\ell}\log^2\bigg(\frac{1-\beta_\ell}{1+\beta_\ell}\bigg) 
  + \frac{1}{\beta_\ell}\text{Li}_2\bigg(\frac{\beta_\ell-1}{\beta_\ell+1}\bigg) \notag \\
  &+ \frac{\pi^2}{12\beta_\ell} + 3 \log \frac{m_\ell}{\mu} + \chi(\mu)\,,
  \label{F_fns}
\end{align}
and
\begin{equation}
 \text{Li}_2(x)=-\int^x_0\diff t\frac{\log(1-t)}{t}\,.
\end{equation}
The contact term $\chi(\mu)$ arises from the counterterm Lagrangian~\cite{Savage:1992ac,GomezDumm:1998gw,Knecht:1999gb}
\begin{align}
  {\cal L}_\text{c.t.} &= \frac{3i\alpha^2}{32\pi^2}(\bar \ell \gamma^\mu\gamma_5 \ell) \Big\{ \chi_1 \text{Tr}\big(Q^2\{U^\dagger,\partial_\mu U\}\big) \notag\\
  &\qquad+ \chi_2 \text{Tr}\big(QU^\dagger Q\partial_\mu U - Q\partial_\mu U^\dagger Q U\big) \Big\}\,,
  \label{L ct}
\end{align}
where $Q=\text{diag}(2/3,-1/3,-1/3)$ is the charge matrix and $\chi(\mu) =-(\chi_1^\text{r}(\mu) + \chi_2^\text{r}(\mu) + 14)/4$ collects the finite parts $\chi_i^\text{r}$ of the LECs.  It is conventional to decompose $\chi$ into long- and short-distance parts
\begin{equation}
  \chi(\mu) = \chi_{\gamma\gamma}(\mu) + \chi^{}_\text{SD}\,,
\end{equation}
where the scale dependence of $\chi_{\gamma\gamma}(\mu)$ compensates that from the term $\sim \log m_\ell / \mu$.  Although the SM prediction for $\chi^{}_\text{SD}$ is known, $\chi_{\gamma\gamma}$ depends on $\chi_{1,2}$ whose values are not fixed by chiral symmetry. However, we can argue as before and observe that if LFU holds, then the SM values of $\chi$ must be equal in both the electron and muon channels. Then, using the chiral realization~\eqref{bosonize} of the $V-A$ current, one obtains an analogous relation to~\eqref{aNP} for the NP Wilson coefficient
\begin{equation}
\label{Wilson_Kll}
  N_K C_{7A}^\text{NP} = -\frac{\alpha}{F_K} \bigg(\frac{2\Gamma_{\gamma\gamma}}{\pi \mk^3}\bigg)^{1/2}  \chi^{}_\text{NP}\,,
\end{equation}
where we have defined $\Gamma_{\gamma\gamma} = \Gamma(K_L\to \gamma\gamma)$, $N_K=G_F V_{ud}V_{us}^*$, and identified $F_0$ with the kaon decay constant $F_K$.  This implies that
\begin{align}
  C_{7A}^{\mu\mu} - C_{7A}^{ee} &=  -\frac{\alpha}{F_K N_K} \bigg(\frac{2\Gamma_{\gamma\gamma}}{\pi \mk^3}\bigg)^{1/2} \big( \chi^{\mu\mu} - \chi^{ee} \big)\notag\\
  &=-4.8\times 10^{-6}\big( \chi^{\mu\mu} - \chi^{ee} \big)\,,
  \label{chi diff}
\end{align}
and thus NP limits can be inferred from precise extractions of $\chi$ in each lepton channel.  Note that although $\chi$ is scale dependent, this dependence drops out in the difference \eqref{chi diff}.  From the measured rates~\eqref{Rellexp} one can use~\eqref{RL} and \eqref{F_fns} to extract $\chi$, up to a twofold ambiguity.  The resulting values for each solution are shown in Table~\ref{tab:chi}, where we see that solution 2 for the electron channel is clearly ruled out.  However, the present data are not precise enough to distinguish among the remaining solutions.

\begin{table}[t]
\renewcommand{\arraystretch}{1.3}
\centering
\begin{tabular}{ccc}\toprule
Channel & $\chi$ (Solution 1) & $\chi$ (Solution 2) \\
\hline
$ee$ & $5.1^{+15.4}_{-10.3}$ & $-\big(57.5^{+15.4}_{-10.3}\big)$ \\
$\mu\mu$ & $3.75\pm0.20$ & $1.52\pm0.20$ \\\botrule
\end{tabular}
\caption{Values of the contact term $\chi(M_\rho)$ extracted from the measured $K_L\to e^+e^-$ and $K_L\to \mu^+\mu^-$ rates.}
\label{tab:chi}
\end{table}

The derivation of~\eqref{chi diff} relies on $\chi$PT$_3$, generalized to include effects due to $\eta$--$\eta'$ mixing.  The leading contribution to the decay is mediated by pseudoscalar poles, $P=\pi^0,\eta,\eta'$, and a constant form factor for the $P\to\gamma^*\gamma^*$ transition. 
At one-loop order, the $P\to\ell^+\ell^-$ decays all involve the same combination of LECs $\chi_{1,2}$  introduced in \eqref{L ct} for $K_L\to\ell^+\ell^-$. In~\cite{Vasko:2011pi,Husek:2014tna} the corresponding $\pi^0\to e^+e^-$ amplitude was calculated, including full radiative corrections.
Compared to Table~\ref{tab:chi}, the resulting extraction $\chi(M_\rho)=4.5\pm 1.0$ from the KTeV measurement~\cite{Abouzaid:2006kk} would favor solution $1$ also for the muon mode.
Moreover, the estimate for two-loop effects based on the double logarithm~\cite{Husek:2014tna}
\begin{align}
 \chi^\text{LL}(M_\rho)&=\frac{1}{36}\bigg(\frac{M_\pi}{4\pi F_\pi}\bigg)^2\bigg(1-\frac{10m_e^2}{\mpi^2}\bigg)\log^2\frac{M_\rho^2}{m_e^2}\notag\\
 &=0.081
\end{align}
indicates that at least for the pion-pole contribution to $K_L\to\ell^+\ell^-$, the one-loop formula should be sufficient.
However, a similar estimate cannot be derived for the $\eta$ channel since at two-loop order, $SU(3)$ breaking effects render the decay amplitude sensitive to $\chi_1-\chi_2$ as well.
An explicit calculation~\cite{Masjuan:2015cjl} for $\eta,\eta'\to\ell^+\ell^-$ based on Canterbury approximants suggests that for those channels, LFUV two-loop effects are indeed significant. 

The potential impact of two-loop corrections has been investigated before in the context of $K_L\to\mu^+\mu^-$ in~\cite{Isidori:2003ts,D'Ambrosio:1997jp}, where large-$N_c$ and chiral arguments suggest that one can replace the (normalized) point-like form factor by the following parametrization 
\begin{align}
f(q_1^2,q_2^2)&=1+\tilde\alpha \bigg(\frac{q_1^2}{q_1^2-M_\rho^2}+\frac{q_2^2}{q_2^2-M_\rho^2}\bigg)\notag\\
&-(1+2\tilde\alpha)\frac{q_1^2q_2^2}{(q_1^2-M_\rho^2)(q_2^2-M_\rho^2)}\,,
\end{align}
where $\tilde\alpha$ is a free parameter.  Based on this parametrization, the $m_\ell$-dependent terms in the $\gamma\gamma$ integral produce a shift in $\chi$ of the form~\cite{Isidori:2003ts}
\begin{equation}
 \Delta\chi(M_\rho)=\frac{\tilde\alpha}{3r_V^2}\bigg[\big(1-10r_\ell^2\big)\log\frac{r_\ell^2}{r_V^2}-\frac{47}{3}r_\ell^2\bigg]-\frac{5r_\ell^2}{3r_V^2}\,,
\end{equation}
which yields $\Delta\chi^{\mu\mu}-\Delta\chi^{ee}=-2.8$, where we have used $\tilde \alpha=-1.69$ as extracted from the slope in $K_L\to\ell^+\ell^-\gamma$~\cite{Cirigliano:2011ny}.
Comparing to the numbers in Table~\ref{tab:chi},
we conclude that once the $ee$ channel can be improved accordingly, additional input from phenomenology, $K_L\to\ell^+\ell^-\gamma$ and $K_L\to\ell^+\ell^-\ell'^+\ell'^-$, will be required to subtract the two-loop corrections and thereby identify potential LFUV contributions.

To illustrate the improvement required in the $ee$ mode for such a test of LFUV in the interesting parameter space, we return to the one-loop relation~\eqref{chi diff} and again
invoke MFV as in~\eqref{C_charged} to translate the kaon-physics limits into the $B$ meson sector%
  \footnote{Using the long-distance amplitude for $K_L\to\mu^+\mu^-$ in~\cite{D'Ambrosio:1997jp,GomezDumm:1998gw}, an upper bound for the short-distance contribution can be obtained. MFV can then be used to extract limits on $C_{10}^{B,\mu\mu}$ directly~\cite{D'Ambrosio:2002ex}.}
\begin{align}
 C_{10}^{B,\mu\mu}-C_{10}^{B,ee}&=\frac{2\pi}{F_K G_F\lambda_t}\bigg(\frac{2\Gamma_{\gamma\gamma}}{\pi \mk^3}\bigg)^{1/2} \big( \chi^{\mu\mu} - \chi^{ee} \big)\notag\\
 &=2.6\bigg(\frac{3.5\times 10^{-4}}{\lambda_t}\bigg)\big( \chi^{\mu\mu} - \chi^{ee} \big)\,.
\end{align}
Suppose the uncertainty in $\Gamma(K_L\to\ell^+\ell^-)$ could be reduced by a factor of $10$, and that the central value remained unchanged. In this case, the second solution for the muon case would be strongly disfavored, given that LFUV if present at all should manifest itself as a small effect, so that $\chi^{\mu\mu} - \chi^{ee}\sim 1.3\pm 1.3$, and, assuming MFV, $C_{10}^{B,\mu\mu}-C_{10}^{B,ee}\sim 3.5\pm 3.5$.
Comparison to~\eqref{C_charged} shows that the sensitivity of thus improved $K_L\to\ell^+\ell^-$ decays to $C_{10}^B$ happens to be similar to the one of a tenfold reduced uncertainty of $K^+\to\pi^+\ell^+\ell^-$ to $C_9^B$. In either case one needs in fact more than an order-of-magnitude improvement to test the $B$-physics anomalies.   

\section{Lepton-flavor-violating decays}
\label{sec:LFV}

\begin{table*}[t]
\renewcommand{\arraystretch}{1.3}
\centering
\begin{tabular}{ccccc}\toprule
 & $K_L\to\mu^\pm e^\mp$ & $K^+\to\pi^+\mu^\pm e^\mp$ & $K_L\to\pi^0\mu^\pm e^\mp$ & $K^+\to\pi^+\mu^\pm e^\mp$ (NA62 projection) \\
\hline
$\big(|C_{7V}^{\mu e}|^2+|C_{7A}^{\mu e}|^2\big)^{1/2}$ & $<1.3\times 10^{-6}$ & $<2.2\times 10^{-5}$ & & $<5.1\times 10^{-6}$\\
$\big(|y_{7V}^{\mu e}|^2+|y_{7A}^{\mu e}|^2\big)^{1/2}$ & & & $<0.040$ &\\
$\big(|C_{9}^{B,\mu e}|^2+|C_{10}^{B,\mu e}|^2\big)^{1/2}$ & $<0.71$ & $<12$ & $<35$ & $<2.7$\\\botrule
\end{tabular}
\caption{Limits on LFV Wilson coefficients from kaon decays. In the case of $K^+\to\pi^+\mu^\pm e^\mp$ only the limit from the channel $K^+\to\pi^+\mu^+ e^-$ is considered. 
The last line shows the corresponding limits in the $B$ system assuming MFV, while the rightmost column refers to the projected limit from NA62~\cite{Ceccucci:CD2015}.}
\label{tab:limits}
\end{table*}

Apart from tiny effects due to neutrino oscillations, LFV does not occur in the SM, so the decay rates can be expressed directly in terms of the NP Wilson coefficients and quark operators based on the chiral realization~\eqref{bosonize}. In general, the decay rate for $K_L\to\ell_1^+\ell_2^-$ takes the form
\begin{align}
 \Gamma\big(K_L\to\ell_1^+\ell_2^-\big)&=(4\pi)^{-1}\mk^3\sqrt{\lambda(1,r_{\ell_1}^2,r_{\ell_2}^2)}F_K^2N^2_K\notag\\
 &\hspace{-60pt}\times\Big\{|C_{7V}^{\ell_1\ell_2}|^2\big(r_{\ell_1}-r_{\ell_2}\big)^2\Big[1-\big(r_{\ell_1}+r_{\ell_2}\big)^2\Big]\notag\\
 &\hspace{-40pt}+|C_{7A}^{\ell_1\ell_2}|^2\big(r_{\ell_1}+r_{\ell_2}\big)^2\Big[1-\big(r_{\ell_1}-r_{\ell_2}\big)^2\Big]\Big\}\,.
 \label{KLll}
\end{align}
In the limit $\ell_1=\ell_2$, the vector component is absent and the expression~\eqref{KLll} reduces to the short-distance part of~\eqref{RL}--\eqref{Wilson_Kll}:
\begin{equation}
 \Gamma\big(K_L\to\ell^+\ell^-\big)=\frac{\mk^3r_\ell^2\beta_\ell}{\pi}|C_{7A}^{\ell\ell}|^2F_K^2N_K^2\,.
\end{equation}
In the context of LFV we need $\ell_1=\mu$ and $\ell_2=e$
\begin{align}
\label{GammaK}
 \Gamma\big(K_L\to\mu^\pm e^\mp\big)&=(4\pi)^{-1}\mk^3r_\mu^2\big(1-r_\mu^2\big)^2F_K^2N_K^2\notag\\
 &\times\big\{|C_{7V}^{\mu e}|^2+|C_{7A}^{\mu e}|^2\big\}\,,\notag\\
 \Br\big[K_L\to\mu^\pm e^\mp\big]&=2.6\big\{|C_{7V}^{\mu e}|^2+|C_{7A}^{\mu e}|^2\big\}\,,
\end{align}
where the mass of the electron has been neglected.

Similarly, we find for the semileptonic decay spectra
\begin{align}
 \frac{\diff\Gamma}{\diff z}\big(K^+\to\pi^+\mu^\pm e^\mp\big) &= \frac{\mk^5N_K^2}{12(4\pi)^3}\big\{|C_{7V}^{\mu e}|^2+|C_{7A}^{\mu e}|^2\big\}\notag\\
 &\hspace{-75pt}\times\sqrt{\bar{\lambda}}\bigg(1-\frac{r_\mu^2}{z}\bigg)^2
 \bigg\{\bar\lambda\bigg(2+\frac{r_\mu^2}{z}\bigg)+3\frac{r_\mu^2}{z}\big(1-r_\pi^2\big)^2\bigg\}\,,\notag\\
 \frac{\diff\Gamma}{\diff z}\big(K_L\to\pi^0\mu^\pm e^\mp\big) &= \frac{\mk^5\tilde N_K^2}{12(4\pi)^3}\big\{|y_{7V}^{\mu e}|^2+|y_{7A}^{\mu e}|^2\big\}\notag\\
 &\hspace{-75pt}\times\sqrt{\bar{\lambda}}\bigg(1-\frac{r_\mu^2}{z}\bigg)^2
 \bigg\{\bar\lambda\bigg(2+\frac{r_\mu^2}{z}\bigg)+3\frac{r_\mu^2}{z}\big(1-r_\pi^2\big)^2\bigg\}\,,
\end{align}
where $r_\mu^2 \leq z \leq (1-r_\pi)^2$, $\tilde N_K=G_F\Im\lambda_t$, and $\mk$ and $\mpi$ denote the charged/neutral particle masses according to each decay. (For simplicity, the $K_{\ell3}$ form factors have been put equal to unity.)
The integrated decay widths are given by
\begin{align}
\label{GammaKpi}
\Gamma\big(K^+\to\pi^+\mu^\pm e^\mp\big)&=\mk^5N_K^2I_+\big\{|C_{7V}^{\mu e}|^2+|C_{7A}^{\mu e}|^2\big\}\,,\notag\\
\Gamma\big(K_L\to\pi^0\mu^\pm e^\mp\big)&=\mk^5\tilde N_K^2I_L\big\{|y_{7V}^{\mu e}|^2+|y_{7A}^{\mu e}|^2\big\}\,,
\end{align}
where the phase space factors are
\begin{equation}
 I_+=7.49\times 10^{-6}\,,\qquad
 I_L=7.99\times 10^{-6}\,,
\end{equation}
so that
\begin{align}
 \Br\big[K^+\to\pi^+\mu^\pm e^\mp\big]&=0.027\big\{|C_{7V}^{\mu e}|^2+|C_{7A}^{\mu e}|^2\big\}\,,\notag\\
 \Br\big[K_L\to\pi^0\mu^\pm e^\mp\big]&=4.7\times10^{-8}\bigg(\frac{\Im\lambda_t}{1.35\times 10^{-4}}\bigg)^2\notag\\
 &\times\big\{|y_{7V}^{\mu e}|^2+|y_{7A}^{\mu e}|^2\big\}\,.
\end{align}

Based on~\eqref{GammaK} and \eqref{GammaKpi}, the experimental limits summarized in Table~\ref{tab:LFV_exp} can be turned into limits on the Wilson coefficients $(|C_{7V}^{\mu e}|^2+|C_{7A}^{\mu e}|^2)^{1/2}$ and $(|y_{7V}^{\mu e}|^2+|y_{7A}^{\mu e}|^2)^{1/2}$. In particular, given that the same combination of Wilson coefficients appears if we neglect the electron mass, the analysis in terms of effective operators allows one to compare the limits from different channels in a model-independent way (this is similar to the analysis of Higgs-mediated LFV in $\mu \to e \gamma$ and $\mu \to e$ conversion in nuclei in~\cite{Crivellin:2014cta}).  The resulting limits are given in the first two lines of Table~\ref{tab:limits}, where the limit on the $C_{7V,7A}$ combination from $K_L\to\mu^\pm e^\mp$ decays is an order of magnitude more stringent than the one from $K^+\to\pi^+\mu^\pm e^\mp$. Even the projected improvement from NA62~\cite{Ceccucci:CD2015} will fall short by a factor of $4$.

As in the case of LFUV, we assume MFV to convert the limits on LFV Wilson coefficients in kaon decays to limits for the $B$-physics coefficients (see~\cite{Lee:2015qra} for a similar analysis). These are shown in the bottom line of Table~\ref{tab:limits}, where in the case of the $K\to\pi\mu e$ decays, the resulting constraints are slightly better than~\eqref{C_charged}, but of similar order of magnitude.  The strongest constraint is obtained from the limit on $K_L\to\mu e$.

\section{Conclusions}
\label{sec:conclusion}

Motivated by the flavor anomalies observed by LHCb in semileptonic $B$ meson decays and CMS/ATLAS in 
$h\to\mu\tau$, we presented an analysis of $K\to\pi\ell^+\ell^-$ and $K\to \ell^+\ell^-$ decays 
to search for lepton flavor (universality) violation in the kaon sector. In general, the search for NP in these decays proves to be very challenging: long-distance contributions from the SM need to be separated from the interesting short-distance effects, both of which enter in poorly known low-energy constants of the $\chi$PT$_3$ expansion.

We observed that in the context of LFUV, this complication is absent if the difference between electron and muon parameters is considered. This simplification is due to the fact that in the SM all interactions (except those involving Higgs-Yukawa couplings) are LFU conserving.  Since the  Higgs corrections are negligible, it follows that the SM decays of kaons to muons or electrons differ only by phase space factors. Thus, any deviation from the SM predictions must be related to LFUV NP which is necessarily short distance once the new particles are assumed to be heavy.

For vector and axial-vector effective operators, we extracted the corresponding limits on the Wilson coefficients of the LFUV operators from $K^+\to\pi^+\ell^+\ell^-$ and $K_L\to \ell^+\ell^-$.  Assuming MFV, we translated the derived limits to the corresponding $B$-physics Wilson coefficients. We found that the kaon limits would need to be improved by at least an order of magnitude in order to probe the parameter space relevant for the explanation of the $B$ meson anomalies and thereby test those anomalies within the MFV hypothesis. 

For the charged $K$-decay, improvements in this direction could be realized at the NA62 experiment, which in our view provides additional motivation to study rare decays besides the main $K^+\to\pi^+\nu\bar\nu$ channel. Constraining LFUV in the neutral decays $K_{L,S} \to \pi^0\ell^+\ell^-$ proves to be challenging, especially since $\Br [K_L\to\pi^0\ell^+\ell^-]$ has not been measured and improved information from the $K_S\to\pi^0\ell^+\ell^-$ spectrum would be required to interpret the $K_L$ branching ratio. The alternative search channel $K_L\to \ell^+\ell^-$ in principle provides access to the axial-vector couplings, but also here improvements by an order of magnitude would be required. The KOTO experiment, mainly motivated by a measurement of $K_L\to\pi^0\nu\bar\nu$, might have the required sensitivity to probe LFUV in the neutral decay if the experiment could be adapted to allow for the detection of the charged leptons in the final state.

Finally, we expressed the decay rates for the LFV decay channels in terms of the corresponding Wilson coefficients and derived the bounds implied by the present experimental limits.
We found that all channels are sensitive to the same combination of Wilson coefficients, with the most stringent bounds presently from $K_L\to\mu^\pm e^\mp$.

We conclude that the upcoming NA62 experiment might have the potential to provide interesting insights into current puzzles in the flavor sector, complementary to direct measurements in $B$ meson decays.  From our analysis, the following scenarios emerge: if NP explanations for the $B$ meson anomalies satisfied MFV, then one should see a signal at the sensitivities discussed in this paper. On the other hand, if the searches at a sensitivity expected from MFV turned out negative or if one saw a signal at current or slightly improved sensitivity, one could immediately infer that any NP scenario explaining the $B$ anomalies would require violations of the MFV hypothesis.

\section*{Acknowledgments}

We thank Evgueni Goudzovski, Hong Ma, and Peter Tru\"ol for helpful communication regarding the NA48/2 and E865 data sets.  We also thank Gerhard Ecker for correspondence on large-$N_c$ realizations of the Gilman--Wise operators, and Karol Kampf, Pere Masjuan, and Pablo Sanchez-Puertas for discussions on the role of $P\to\ell^+\ell^-$ decays in $K_L\to\ell^+\ell^-$.  L.C.T.\ thanks Ding Yu Shao and Xavier Garcia i Tormo for useful discussions.  
Financial support by MIUR under the project number 2010YJ2NYW,
the DOE (Grant No.\ DE-FG02-00ER41132),
and the Swiss National Science Foundation
is gratefully acknowledged.
A.C.\ is supported by a Marie Curie Intra-European Fellowship of the European Community's 7th Framework Programme (contract number PIEF-GA-2012-326948).

\bibliography{BIB}      

\end{document}